\documentclass[final,5p,times,twocolumn]{elsarticle}

\usepackage{amssymb}
\usepackage{amsmath}
\usepackage{xcolor}
\usepackage{float}
\usepackage{xcolor}
\usepackage{graphicx}
\usepackage{soul}

\usepackage{hyperref}

\hypersetup{
    colorlinks=true, 
    linktoc=all,     
    citecolor=black,
    filecolor=black,
    linkcolor=black,
    urlcolor=black
}

\biboptions{comma,square,compress}

\newcounter{bla}

\journal{Computer Physics Communications}

\DeclareUnicodeCharacter{2060}{}
\DeclareUnicodeCharacter{0302}{}

\usepackage{xr}

\makeatletter

\newcommand*{\addFileDependency}[1]{
  \typeout{(#1)}
  \@addtofilelist{#1}
  \IfFileExists{#1}{}{\typeout{No file #1.}}
}
\makeatother

\newcommand*{\myexternaldocument}[1]{%
    \externaldocument{#1}%
    \addFileDependency{#1.tex}%
    \addFileDependency{#1.aux}%
}

\myexternaldocument{TMM-Sim-SI}

\begin{document}

\begin{frontmatter}

\title{TMM$-$Sim: A Versatile Tool for Optical Simulation of Thin$-$Film Solar Cells}

\author[a]{Leandro Benatto\corref{author}}
\author[a]{Omar Mesquita}
\author[b]{Kaike R. M. Pacheco}
\author[b]{Lucimara S. Roman}
\author[b]{Marlus Koehler}
\author[a,c]{Rodrigo B. Capaz}
\author[a]{Grazi\^{a}ni Candiotto\corref{author}}

\address[a]{Institute of  Physics, Federal University of Rio de Janeiro, 21941$-$909, Rio de Janeiro$-$RJ, Brazil.}
\address[b]{Department of Physics, Federal University of Paran\'{a}, 81531$-$980, Curitiba$-$PR, Brazil.}
\address[c]{Brazilian Nanotechnology National Laboratory (LNNano), Brazilian Center for Research in Energy and Materials (CNPEM), 13083$-$100, Campinas$-$SP, Brazil.}

\cortext[author]{Corresponding authors: \\ lb08@fisica.ufpr.br and gcandiotto@iq.ufrj.br}

\begin{abstract}
The Transfer Matrix Method (TMM) has become a prominent tool for the optical simulation of thin$-$film solar cells, particularly among researchers specializing in organic semiconductors and perovskite materials. As the commercial viability of these solar cells continues to advance, driven by rapid developments in materials and production processes, the importance of optical simulation has grown significantly. By leveraging optical simulation, researchers can gain profound insights into photovoltaic phenomena, empowering the implementation of device optimization strategies to achieve enhanced performance. However, existing TMM$-$based packages exhibit limitations, such as requiring programming expertise, licensing fees, or lack of support for bilayer device simulation. In response to these gaps and challenges, we present the TMM Simulator (TMM$-$Sim), an intuitive and user$-$friendly tool to calculate essential photovoltaic parameters, including the optical electric field profile, exciton generation profile, fraction of light absorbed per layer, photocurrent, external quantum efficiency, internal quantum efficiency, and parasitic losses. An additional advantage of TMM$-$Sim lies in its capacity to generate outcomes suitable as input parameters for electro$-$optical device simulations. In this work, we offer a comprehensive guide, outlining a step$-$by$-$step process to use TMM$-$Sim, and provide a thorough analysis of the results. TMM$-$Sim is freely available, accessible through our web server (\href{https://nanocalc.org/}{nanocalc.org}), or downloadable from the \href{https://github.com/NanoCalc/TMM-Sim}{TMM$-$Sim} repository (for \textit{Unix}, \textit{Windows}, and \textit{macOS}) on \textit{GitHub}. With its user$-$friendly interface and powerful capabilities, TMM$-$Sim aims to facilitate and accelerate research in thin$-$film solar cells, fostering advancements in renewable energy technologies.
\end{abstract}


\end{frontmatter}

\section{Introduction}
Solar energy is widely recognized as a promising solution to tackle the pressing environmental and energy challenges facing the planet. In this regard, solution$-$processed thin$-$film photovoltaic devices have emerged as an innovative and cost$-$effective alternative to address these issues \cite{chaturvedi2021all}. These devices offer the advantage of being printed layer by layer, and they possess interesting features such as lightness, flexibility, and semi$-$transparency \cite{wang2021recent,barreto2023improved,spada2013role}. The unique characteristics of these devices are largely dependent on the materials used and the thickness of the deposited layers\cite{benatto2024plq}.

Over the years, significant academic and industrial efforts have been dedicated in optimizing these thin$-$film photovoltaic devices \cite{duan2020progress, li2022recent,candiotto2017,candiotto2024exploring,souza2022binding}. Likewise, the chemical structure of the active layer materials have experienced a substantial progress, leading to notable enhancements in device efficiency \cite{wadsworth2019critical,zheng2020pbdb,meredith2020nonfullerene,zhao2020chlorination,benatto2023enhancing,benatto2021conditions}. In order to enhance the ecological sustainability of these devices, various studies have focused on replacing toxic solvents with environmentally$-$friendly alternatives in the production processes \cite{wouk2018charge,de2020effects,wan2022all,li2023layer}. This approach aims to reduce the negative impact of manufacturing processes on the environment and improve the sustainability of the devices. Recently, the use of these kind of photovoltaic devices has expanded from outdoor to indoor applications, with a recent emphasis on improving its performance for the latter \cite{cui2019wide,miranda2021efficient}. Given that  these devices are becoming commercially viable \cite{karki2021path,rong2018challenges}, it is imperative to explore multiple optimization strategies to be applied over the next few years.

The theoretical modeling of devices has proven to be a valuable approach to shift certain paradigms that have arisen during the maturation of this emerging technology \cite{coropceanu2007charge,oberhofer2017charge,merces2021,candiotto2020,benatto2023fret}. One  of the most significant development models for the area was the Transfer Matrix Method (TMM) \cite{inganas2018organic}. The TMM was initially developed in 1999 \cite{pettersson1999modeling} and has since then emerged as a potent and multifaceted tool for modeling the  optical properties of thin film systems, including solar cells and other optoelectronic devices. In essence, the TMM simulates the electromagnetic wave propagation in each layer of a multilayered optical structure using transfer matrices that considers factors such as transmission, reflection, and phase changes. Thus  crucial information of the device such as the optical electric field profile ($|E(x)|^{2}$), exciton generation profile ($Q(x)$), fraction of light absorbed per layer ($A(x)$), photocurrent ($J_{photo}$), external quantum efficiency ($EQE$, also know as $IPCE$, incident photon to current collection efficiency), internal quantum efficiency ($IQE$), and the parasitic losses can be accessed by using the TMM formalism. Additionally, it  has been demonstrated that this method  is a valuable tool to optimize the solar cell's performance, understand light absorption and exciton generation, and identify opportunities for enhancing device efficiency \cite{peumans2003small}. With its exceptional versatility, the TMM has gained significant popularity among worldwide researchers focused on organic \cite{gevaerts2011discriminating,zang2018effect,benatto2020comparing,gavim2022modelling} and perovskite$-$based \cite{ball2015optical,van2017optical, lin2015electro} solar cells.

In order to use the TMM, it is first necessary to specify the wavelength ($\lambda$) dependent complex refractive index $\eta(\lambda)$ (optical constants) and thickness of each layer in the system. The complex index of refraction is composed by a real (refractive index) and an imaginary (extinction coefficient) parts, $\tilde{n}(\lambda)=\eta(\lambda)+i\kappa(\lambda)$\cite{benatto2024ricalc}. The refractive index is responsible for describing the propagation of light, while the extinction coefficient describes the interaction of light with the material \cite{singh2002refractive}.

\begin{figure}[!t]
\centering
    \includegraphics[width=\linewidth]{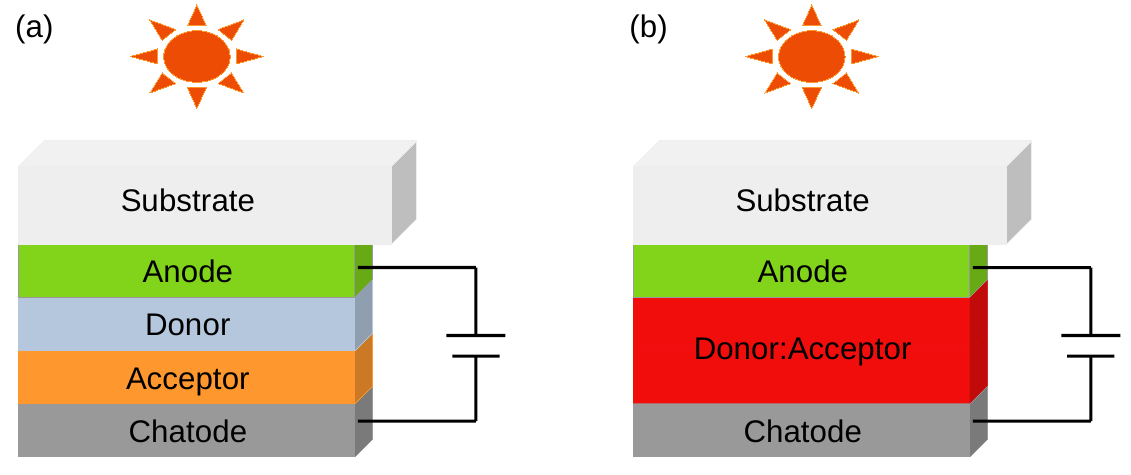}
    \caption{Organic solar cells: (a) bilayer structure and (b) BHJ structure.}
    \label{fig-structures}
\end{figure}

In this manuscript we skip to describe how to obtain $|E(x)|^2$ using TMM since those procedures are well documented in textbooks \cite{sun2017organic} (and in the cited references) but rather we concentrate on demonstrate the application of our software  Transfer Matrix Method Simulator (TMM$-$Sim). This simulation tool is specifically designed to model devices that feature up to 10 stacked layers. Moreover, it offers the capability to simulate two distinct device architectures: the bilayer and bulk heterojunction (BHJ) structures \cite{ostroverkhova2016organic}, see Figure \ref{fig-structures}. In the bilayer structure, the active layer consists of stacked electron donor (D) and electron acceptor (A) materials. Conversely, the binary BHJ structure features two active  materials (D and A) that are blended throughout the film thickness. Ternary BHJ structures involve the use of three materials mixed in the active layer. Previous research has  shown that this particular approach effectively increases the efficiency of the device by enhancing key parameters such as light harvesting, photocurrent generation, and others that are critical for optimal performance \cite{wen2022efficient,guguloth2021improved,cao2021over,zeng2022enhanced}, thereby making it a valuable contribution to the field. Despite intrinsically composed of two or more materials, the BHJ structure is typically treated as a single optically  homogeneous media. The only variation considered is in the spectrum of the refractive index applied in the simulation. Therefore, the program can also be used to simulate solar cells made of a single photovoltaic material, which have a simpler structure and a more stable morphology of the active layer with a lower energy disorder, resulting in a higher upper limit for the efficiency and stability of the cells \cite{price2022free,zhang2022single}.

There are already some packages that utilize the TMM. However, some of them are not intuitive for the ordinary user since they require a certain level of programming expertise \cite{burkhard2010accounting,McGehee}, while others come with a costly licensing fee \cite{SETFOS}. We would like to highlight that current free calculation packages neglect the comprehensive implementation of device simulation for a bilayer structure. This is a crucial lack of capability given that the bilayer structures serve as a useful tool for studying the compatibility of active layer materials and determining the diffusion length of excitons \cite{ran2017impact,firdaus2020long,park2021photophysical}. These factors have motivated us to develop the TMM$-$Sim, a user$-$friendly software package that is freely accessible to simulate bilayer and BHJ devices. The TMM$-$Sim is a simple and powerful tool providing a graphical interface to facilitate its use. 

\section{Software architecture, implementation and requirements}
TMM$-$Sim is a license free code that offers free access to users. It can be utilized conveniently through a dedicated web server (\href{https://nanocalc.org/}{nanocalc.org}) or by downloading the binary files compatible with \textit{Unix}, \textit{Windows}, and \textit{macOS} operating systems, which are available at the \href{https://github.com/NanoCalc/TMM-Sim}{TMM$-$Sim} repository on \textit{GitHub}. This easy accessibility ensures researchers and users from various platforms can easily employ TMM$-$Sim to perform their simulations efficiently. The TMM$-$Sim software is written in Python 3 (v. 3.10) \cite{van2009} and employs four Python libraries: \textit{Pandas} \cite{pandas2010}, \textit{NumPy} \cite{harris2020}, ⁠\textit{SciPy} \cite{virtanen2020} and \textit{Matplotlib} \cite{hunter2007}⁠. The software was designed to be user$-$friendly and take up little disk space, around 80MB.

\section{Program and Application}
Before beginning the program demonstration, it is crucial to emphasize that the calculation method implemented here has been thoroughly validated. Specifically, we conducted a rigorous comparison of our BHJ device results with simulations obtained from a well$-$established package \cite{burkhard2010accounting,McGehee}. The results of this comparative analysis, presented in Figure S1, demonstrate perfect alignment between our implementation and those generated by the established simulation package. This confirms the accuracy and reliability of our methodology.

\subsection{BHJ device simulation}\label{sec:KPQ}
\begin{figure}[!t]
\centering
    \includegraphics[width=\linewidth]{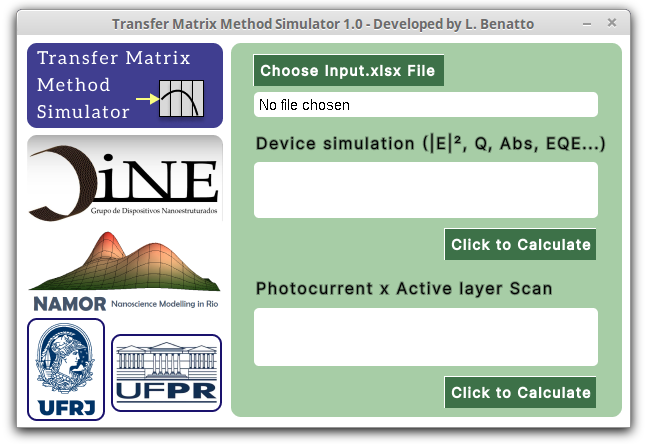}
    \caption{Main interface.}
    \label{fig-interface}
\end{figure}

\begin{figure}[!t]
\centering
    \includegraphics[width=\linewidth]{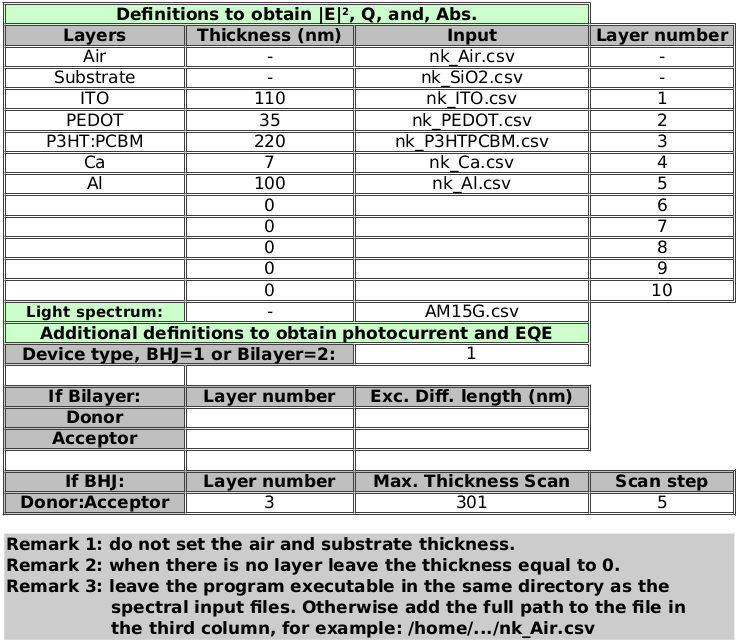}
    \caption{Input for BHJ system.}
    \label{fig-input-bhj}
\end{figure}

The program interface shown in Figure \ref{fig-interface} has a button called "Choose Input.xlsx File" where the user must add the input file with the information necessary to carry out the simulation. All the information about the device can be easily filled in this file, as can be seen in Figure \ref{fig-input-bhj} for a device with the BHJ structure. In the first, second, and third columns the name, thickness, and refractive index data of the layer can be filled. It is important to note that the instructions appended at the end of the input file provide guidance to the user on how to complete it. 

As can be seen in Figure \ref{fig-input-bhj}, the incident light spectrum can be also specified. To complete the input file, four additional specifications are required. 1) It is necessary to set the device type. 2) The corresponding active layer number needs to be defined. 3) The maximum thickness of the active layer must be stated. This information is used to calculate the photocurrent as a function of layer thickness. 4) The scan step should also be included. If the device is a bilayer (as will be demonstrated in the next section), the exciton diffusion length in the donor and acceptor layers must also be supplied. Upon correctly filling the input file and inserting it into the program, several quantities relevant to the photovoltaic process can be computed.

\begin{figure*}[t!]
\centering
    \includegraphics[width=\linewidth]{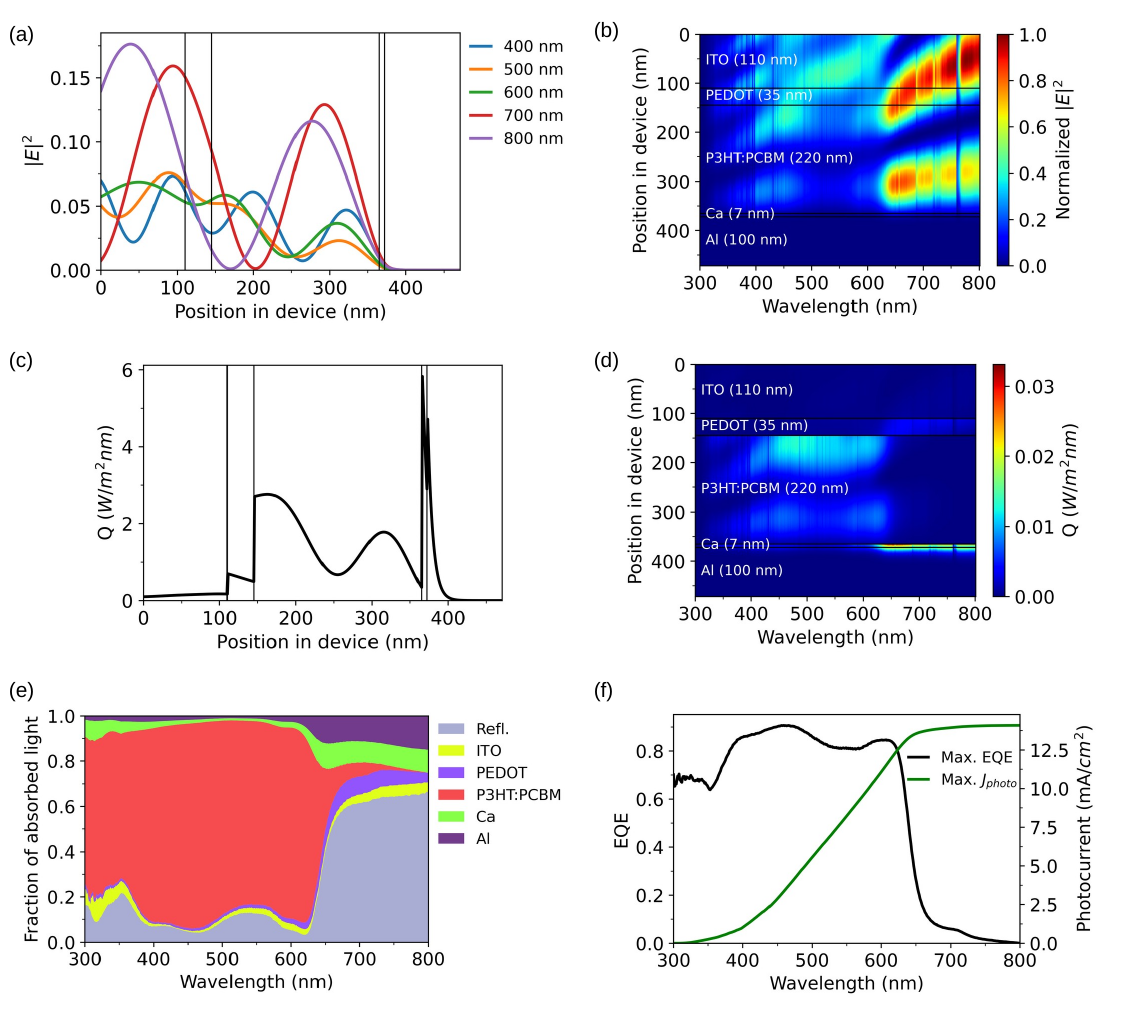}
    \caption{Results for BHJ device simulation, glass$/$ITO(110 nm)$/$PEDOT(35 nm)$/$P3HT:PCBM(220 nm)$/$Ca(7 nm)$/$Al(100 nm). (a) Optical electric field profile. (b) Heatmap of $|E(x)|^{2}$. (c) Exciton generation profile. (d) Heatmap of $Q_{j}(x)$. (e) Absorption per layer. (f) EQE and integrated photocurrent.}
    \label{fig-results-bhj}
\end{figure*}

The input example presented here is a well$-$known organic photovoltaic device in the BHJ structure. The complete structure of the device is glass$/$ITO(110 nm)$/$PEDOT(35 nm)$/$P3HT:PCBM(220 nm)$/$Ca(7 nm)$/$Al(100 nm). In this BHJ structure, ITO (indium tin oxide) is the transparent anode electrode, PEDOT (poly(3,4$-$ethylenedioxythiophene)) is a conducting polymer acting as hole transport$/$electron block layer (HTL$/$EBL), P3HT (poly(3$-$hexylthiophene)) is p$-$type organic semiconducting (electron donor polymer), PCBM ([6,6]$-$phenyl$-$C61$-$butyric acid methyl ester) is a n$-$type organic semiconductor (electron acceptor molecule), Ca (calcium) acts as electron transport$/$hole block layer (ETL/HBL) and Al (aluminium) is the cathode electrode. The refractive index of the materials was obtained from the experimental data presented in refs \cite{burkhard2010accounting,McGehee} and its graphical representation can be seen in Figure S2 of the supplementary information. In the optical simulation of the device, due to the large thickness ($\sim$ 1 mm) and nonuniformity in the thickness of the glass, as well as the finite bandwidth of the light source, the transmission of a beam of light through the glass must be treated as being incoherent with respect to other beams \cite{pettersson1999modeling}. Considering that user$-$supplied refractive index spectra may cover different ranges of the light spectrum, the program automatically identifies the range of wavelengths where all spectra overlap and performs calculations within that range. At the end of the device simulation, the program provides the text files with the results and generates their graphical representation, as can be seen in Figure \ref{fig-results-bhj}.

The first two results provided by the program are the profile of the optical electric field ($|E(x)|^2$, Figure \ref{fig-results-bhj}a) for some wavelengths and the $|E(x)|^2$ heatmap for all considered spectrum (Figure \ref{fig-results-bhj}b). These two outputs provide the pattern of maximums and minimums of the optical field for the combination of materials and thicknesses considered.

Another important result obtained is the exciton generation profile for each layer $j$ given by \cite{sun2017organic},

\begin{equation}\label{eq-Q}
Q_{j}(x)=\dfrac{2\pi c \epsilon_{0}\kappa_{j}\eta_{j}|E(x)|^2}{\lambda},
\end{equation}

\noindent where $x$ is the distance in nm, $c$ is speed of light in vacuum, $\epsilon_{0}$ is permittivity of vacuum, and $\lambda$ is the vacuum wavelength.  The unit of $Q$ is W. m$^{-2}$. nm$^{-1}$. The calculation of $Q_{j}(x)$ indicates where energy is being dissipated within the device. For a good operation of the device, it is important that the points of maximum optical field  are restricted to the limits of the active layer where exciton generation is possible Energy dissipation outside the active layer region must be avoided to decrease the light filtering effect (parasitic absorption losses). Figure \ref{fig-results-bhj}c shows the exciton generation profile integrated for all considered spectrum. Furthermore, the program concomitantly generates a heatmap  that gives the magnitude of $Q_{j}(x)$ for each wavelength, see Figure \ref{fig-results-bhj}d. The discontinuity observed at the layer interfaces is attributed to the abrupt variations in the values of $\kappa$ and $\eta$, due to the transition between materials. The exciton generation profile inside the active layer is a fundamental parameter for electro$-$optical device simulations using numerical drift$-$diffusion (DD) model \cite{li2021organic,nyman2021requirements}. Note that the energy dissipated in the active layer has two maximums, one closer to the interface with PEDOT and another closer to the interface with Ca. For wavelengths above 650 nm, the energy dissipation in the active layer drops sharply due to the low absorption coefficient of the materials in this region. This effect also appears clearly in the following results of light absorption fraction per layer and external quantum efficiency (see Figures \ref{fig-results-bhj}e and \ref{fig-results-bhj}f).

From the simulation of the optical process in the device it is also possible to obtain absorbed fraction of incident light for each layer,

\begin{equation}\label{eq-A}
A_{j} = \dfrac{1}{S_{0}} \int^{dj} Q_{j}(x)dx,
\end{equation}

\noindent where $S_{0}$ is the irradiance from air for a given wavelength. This information is crucial for a variety of applications, such as the comparison among devices that utilize different materials outside the active layer region or even devices using alternative materials within this layer \cite{benatto2020comparing}. Furthermore, the program is capable of computing both the fraction of light reflected by the device and, in the case of a semi$-$transparent device, the fraction of light transmitted through the device.

The program also calculates the maximum external quantum efficiency and maximum photocurrent generated by the BHJ device. The equation that defines $EQE$ is given by \cite{pettersson1999modeling},

\begin{equation}\label{eq-EQE}
EQE (\%) = 1240 \dfrac{J_{Photo}}{\lambda S_{0}},
\end{equation}

where $J_{Photo}$ is the  photocurrent (mA$/$cm$^{2}$) under short$-$circuit condition \cite{gavim2022modelling,khanam2019modeling}. In the simulation of the BHJ device, it is assumed that each photon absorbed in the active layer contributes to the photocurrent, resulting in an internal quantum efficiency of 100\%. Therefore, the percentage of light absorbed in the active layer of the BHJ device will be equal to the external quantum efficiency. The simulated external quantum efficiency can be compared with experimental results to gauge how far from the ideal behavior a determined  device is which gives an idea of how much it can still be optimized.

The TMM$-$Sim packaged  also precisely determines the parasitic absorption losses caused by coherence and absorption in the non$-$photoactive layers \cite{ball2015optical,van2017optical}. A table with the photocurrent losses by reflection, transmission or absorption in the non$-$photoactive layers is provided allowing a deep analysis of the simulated device. The investigated parasitic losses within the selected wavelength range found for the BHJ device under consideration are: 8.09 mA$/$cm$^{2}$ for reflection, 0.75 mA$/$cm$^{2}$ for ITO, 1.06 mA$/$cm$^{2}$ for PEDOT, 1.51 mA$/$cm$^{2}$ for Ca and 1.79 mA$/$cm$^{2}$ for Al.

\begin{figure}[!t]
\centering
    \includegraphics[width=\linewidth]{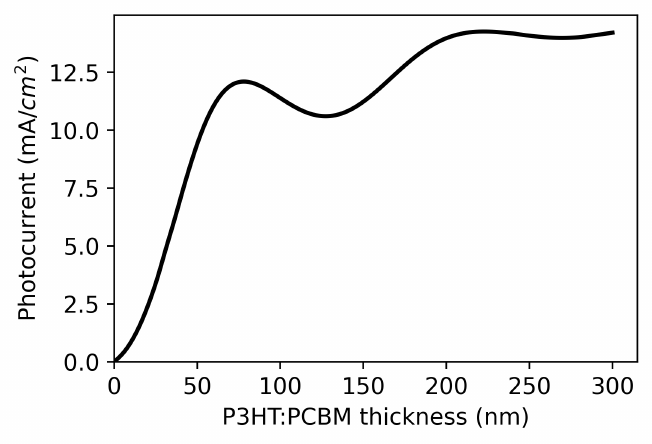}
    \caption{Photocurrent versus active layer thickness for the BHJ device.}
    \label{fig-bhj-scan}
\end{figure}

A second simulation option involves calculating the photocurrent versus the thickness of the active layer. This information is useful because it can estimate the optimal active layer's thickness for a determined BHJ configuration before the actual fabrication of the device. For this calculation, the user must define a maximum thickness and the simulation's step size as input (see Figure \ref{fig-input-bhj}). Given these data, the calculation is performed by pressing the calculation button below the words "Photocurrent x Active layer Scan". If the step size is greater than 1 nm, the program will interpolate the results to provide the photocurrent for all thicknesses within the analyzed range (using a step$-$size of 1 nm). The Figure \ref{fig-bhj-scan} depicts the outcome of the BHJ device simulated in this study. The first photocurrent maximum is obtained for a thickness of 79 nm while the second maximum for 223 nm. It is essential to recognize that as the thickness of the active layer increases, there is a higher probability of charge carrier recombination during the transportation process towards the collection electrodes. Figure \ref{fig-bhj-scan} serves as an exemplary representation of an ideal scenario where no recombination of charge carriers occurs during the transport process. For active layer materials that have a thickness$-$dependent refractive index spectrum, the photocurrent versus active layer thickness calculation should be avoided, as it will not provide results consistent with the reality of the system.

To conclude this section focused on simulating a BHJ device, we will now explore the simulation of a high$-$efficiency device that has emerged as a benchmark for organic solar cells \cite{yuan2019single,guo2021recent,shoaee2023we}. This device incorporates the PM6 polymer along with the non-fullerene acceptor molecule Y6 within its active layer, presenting the following structure in the inverted form glass$/$ITO$/$ZnO$/$PM6:Y6$/$MoO$_3/$Ag. The inverted device has better long$-$term ambient stability. To conduct the simulation, we employed the refractive index of the PM6:Y6 blend as obtained from reference \cite{kerremans2020optical}. The refractive indices of all materials comprising the device are provided in Figure S3 of the supplementary information. The simulation results were presented in Figure S4 and show the EQE spectrum around 80\% in a wide range of the spectrum and photocurrent around 25 mA/cm$^2$. These findings are closely aligned with the literature. The photocurrent as a function of the active layer thickness is illustrated in Figure S5, exhibiting analogous behavior to that predicted theoretically and observed experimentally by Li $et$ $al.$ \cite{li2021organic}.

\subsection{Bilayer device simulation}\label{sec:KPQ}
\begin{figure}[!t]
\centering
    \includegraphics[width=\linewidth]{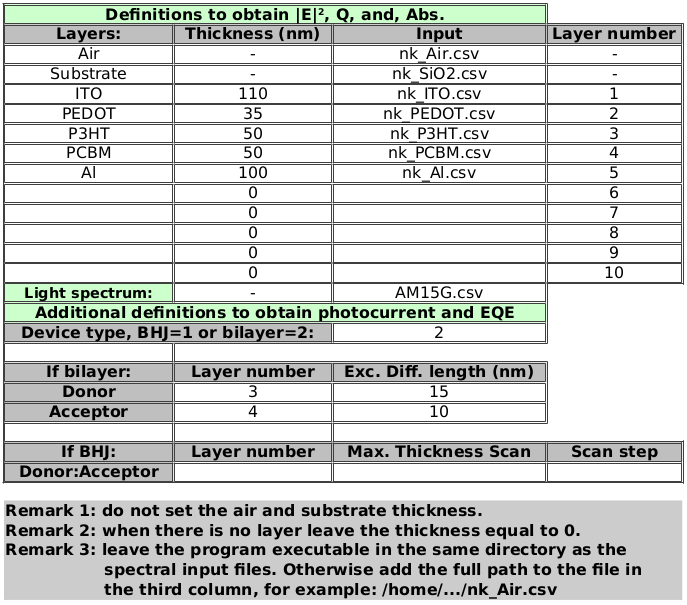}
    \caption{Input for bilayer system.}
    \label{fig-input-bilayer}
\end{figure}

Next, we will proceed to demonstrate the simulation of the bilayer device. This simulation requires defining the exciton diffusion length in each material within the active layer. Figure \ref{fig-input-bilayer} exemplifies an input file filled for device simulation. Note that now the device type is defined such as 2, which indicates a bilayer structure mode. We will demonstrate the programs' capabilities to simulate bilayer devices assuming the structure glass$/$ITO(110 nm)$/$PEDOT(35 nm)$/$P3HT(50 nm)$/$PCBM(50 nm)$/$Al(100 nm).

\begin{figure*}[!t]
\centering
    \includegraphics[width=\linewidth]{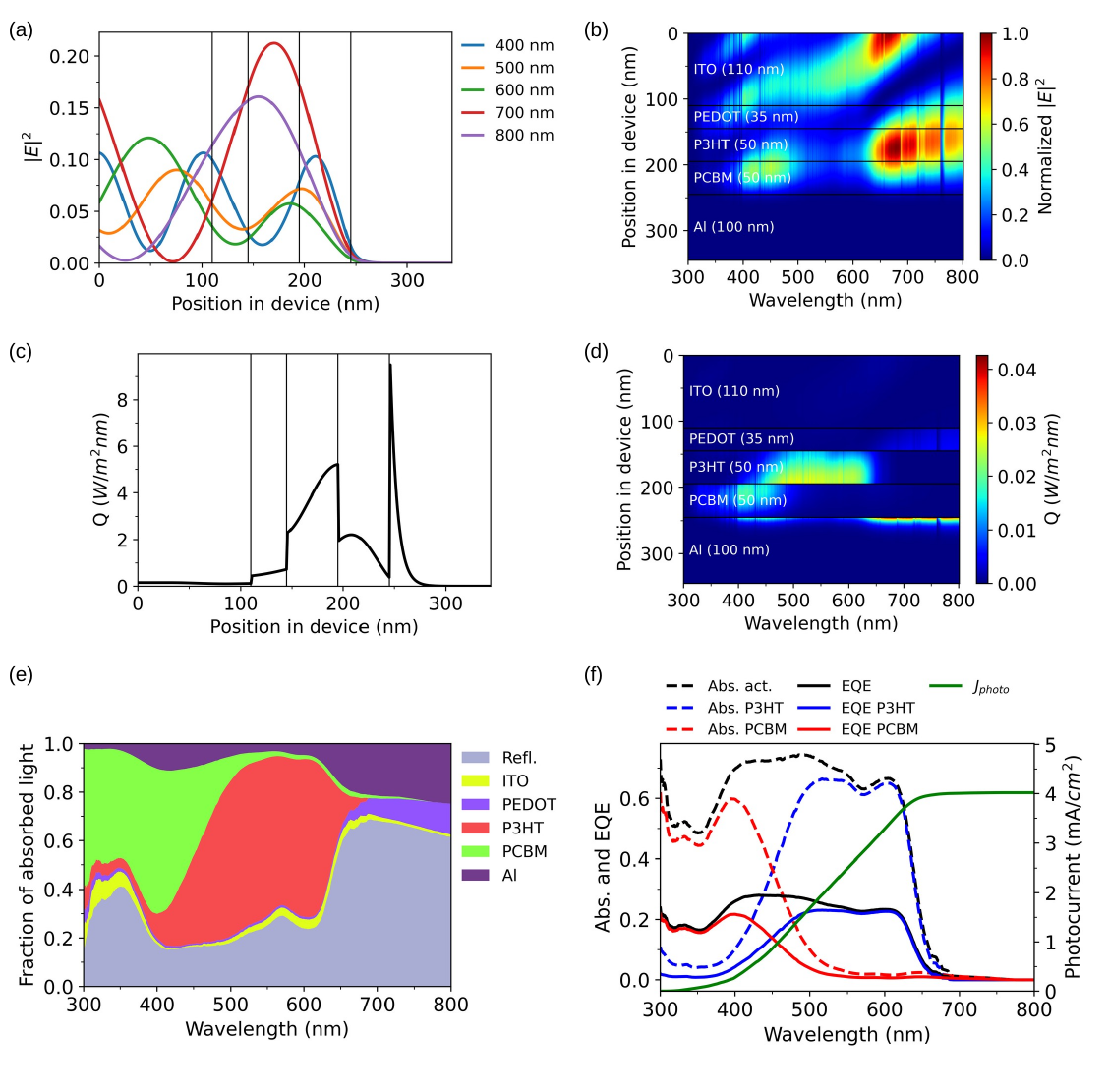}
    \caption{Results for bilayer device simulation, glass$/$ITO(110 nm)$/$PEDOT(35 nm)$/$P3HT(50 nm)$/$PCBM(50 nm)$/$Al(100 nm). (a) Optical electric field profile. (b) Heatmap of $|E(x)|^{2}$. (c) Exciton generation profile. (d) Heatmap of $Q_{j}(x)$. (e) Absorption per layer. (f) EQE, integrated photocurrent, and active layer absorption.}
    \label{fig-results-bilayer}
\end{figure*}

The calculation procedure for simulating certain quantities such as $|E(x)|^2$, $Q(x)$ and absorption per layer is identical for both BHJ and bilayer devices. Nevertheless, the presence of a sharp and well$-$defined donor$/$acceptor (D$/$A) interface, as evidenced by the abrupt variation of the exciton generation profile in Figure \ref{fig-results-bilayer}c, characteristic of a bilayer architecture, a notable difference arises in the calculation of $EQE$: it now depends on the exciton diffusion length ($L$). This parameter is utilized to solve the exciton diffusion equation at steady state \cite{pettersson1999modeling}

\begin{equation}\label{eq-dif}
\dfrac{d^{2}n(x)}{dx^{2}} = \beta^{2}n(x)-\dfrac{Q_{j}(x)}{D}\dfrac{1}{h\nu},
\end{equation}

\noindent where $n$ is the exciton density, $D$ is the diffusion constant, $h \nu$ is the excitation energy of the incident light, and $\beta$ is the reciprocal of the diffusion length ($\beta=1/L=1/\sqrt{\tau D}$, where $\tau$ is the mean lifetime of the exciton). The Eq. \ref{eq-dif} assumes that the generation of excitons occurs with 100\% quantum efficiency, implying that each photon gives rise to an exciton. To obtain a solution for Eq. \ref{eq-dif} for donor and acceptor layers, numerical methods can be employed under the assumption that the interfaces of the active layer function as ideal exciton sinks. Specifically, all excitons are able to either dissociate into free charges or recombine at the interfaces, thereby resulting in boundary conditions given by $n=0$ at $x=0$ (left interface) and $x=d$ (right interface). The short$-$circuit exciton current density at the interface $x=0$ and $x=d$ can be found as

\begin{equation}\label{eq-jexc}
J_{exc} = D\dfrac{dn}{dx}\bigg|_{x=0} \quad \textrm{and} \quad J_{exc} = - D\dfrac{dn}{dx}\bigg|_{x=d},
\end{equation}

\noindent which is related to the short$-$circuit photocurrent through $J_{photo}=q \theta J_{Exc}$, where $q$ is the electronic charge and $\theta$ is the exciton dissociation efficiency at the interface (assumed  100\%). The exciton dissociation in the acceptor layer can happen at both interfaces, D$/$A ($x=0$) and A$/$Al ($x=d$), where the offset between the frontiers orbitals at the D$/$A interface provides the driving force necessary to dissociate the excited state. On the other hand, note that the exciton dissociation in the donor layer happens only at the D$/$A interface.

In some cases, bilayer devices have longer active layer thickness than the exciton diffusion length, leading to increased recombination losses. As a result, it is crucial to adjust the active layer thickness of bilayer devices in accordance with the exciton diffusion length in the donor and acceptor materials. The exciton diffusion length in organic semiconductors is known to depend on the morphology of the formed film. Estimating this length can be done using various approaches and this topic is subject of a great debate in the literature \cite{belova2022effect,firdaus2020long}. Here, to demonstrate the program, we will assume the exciton diffusion lengths of 15 nm for P3HT and 10 nm for PCBM \cite{tamai2015exciton}.

The results obtained with the simulation of the bilayer device are shown in Figure \ref{fig-results-bilayer}. The profile of the optical electric field in Figure \ref{fig-results-bilayer}a reveals that wavelengths of 400 nm, 500 nm, and 600 nm exhibit local maxima near the interface  between donor and acceptor materials, which is important to maximize the generation of excitons idue to the high dissociation power of this region. The presence of those maximums are confirmed by the heatmap of $|E(x)|^2$ in Figure \ref{fig-results-bilayer}b and the results of $Q(x)$ in Figures \ref{fig-results-bilayer}c and \ref{fig-results-bilayer}d. Interesting to note that, unlike the BHJ device, the simpler feature of bilayers with distinct regions of D or A  species  enables to quantify  the  fraction of light absorbed and the $EQE$ contribution for each material composing the active layer. Due to the shorter exciton diffusion length relative to the layer's thickness,  only a fraction of the generated excitons end up being converted into  in bilayer devices. This feature decreases the $EQE$  (see the results of Abs, $EQE$ and Integrated Photocurrent in Figures \ref{fig-results-bilayer}e and \ref{fig-results-bilayer}f). The disparity between the amount of light absorbed by the active layer and the $EQE$ is the amount of photons that do not contribute to the photocurrent. It is important to point out that, by comparing the measured $EQE$ with the calculated one, it is possible to estimate the diffusion length of the excitons and to study the intermixing at the D$/$A interface \cite{gevaerts2011discriminating,benatto2020comparing}.

\begin{figure}[!t]
\centering
    \includegraphics[width=\linewidth]{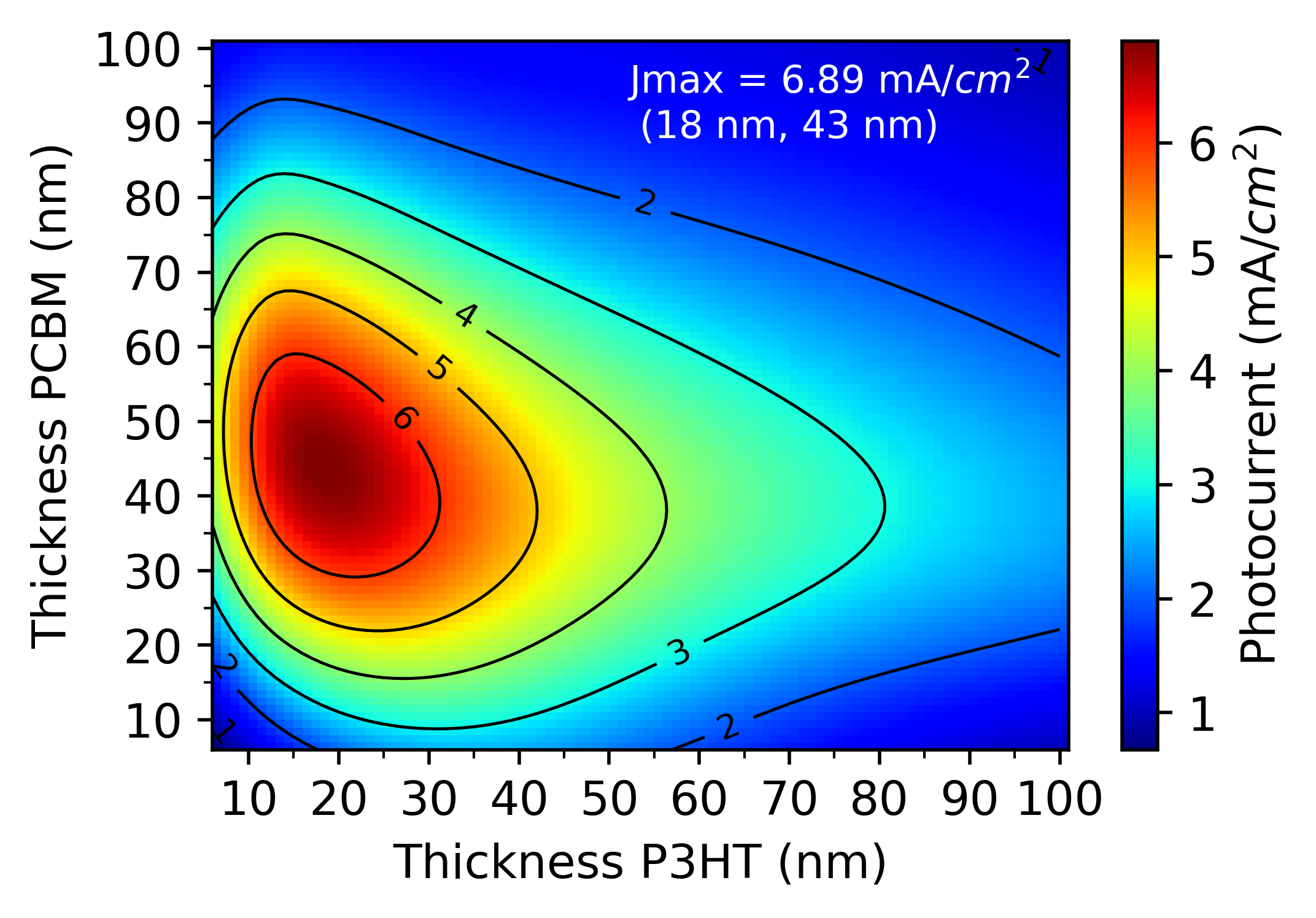}
    \caption{Photocurrent versus active layer thickness for the bilayer device.}
    \label{fig-bilayer-scan}
\end{figure}

To enhance both the photocurrent and $EQE$ of the bilayer device, it is essential to optimize the thickness of the two materials comprising its active layer. The TMM$-$Sim software facilitates this calculation by incorporating a dedicated feature in its graphical interface, accessible through the second button. Once the calculation is finished, the program generates a color graph with contour lines that displays the optimal thickness combination, which maximizes the photocurrent of the device, as depicted in Figure \ref{fig-bilayer-scan}. By comparing the performance of the initial device (50 nm, 50 nm) with the optimized one (18 nm, 43 nm), a significant improvement in photocurrent was achieved, resulting in an increase from 4.02 mA$/$cm$^{2}$ to 6.89 mA$/$cm$^{2}$. This variation corresponds to a photocurrent  increase of 58\%.

\section{Conclusions}
In summary, this work demonstrate  the capabilities of the TMM$-$Sim program to simulate optical processes and performance features of thin$-$film photovoltaic devices. The application of the program to typical BHJ and bilayer structures demonstrates its usefulness and simple handling due to a easy$-$to$-$fill input file. Several results were obtained according to the type of simulated device (BHJ or bilayer). This personalized way of simulating the devices expands the application range of the program. In addition, the package incorporates an extra functionality that enables to estimate the variation of the photocurrent as a function of the active layer thickness. This feature permits to anticipate the optimum active layer thickness before the device fabrication.  It is important to highlight that all graphic results generated by the program are accompanied by data files that the user can utilize to create their own figures or perform further analysis. Moreover, it is noteworthy that the program can also be used to accurately simulate state$-$of$-$the$-$art single$-$junction devices with high efficiency, including those with an inverted structure such as the PM6:Y6 system. We point out that since this is the first version of the program, suggestions will be welcome for its future improvement and launch of new updated versions. 

\section*{Declaration of Competing Interest}
\noindent The authors declare that they have no known competing financial interests or personal relationships that could have appeared to influence the work reported in this paper.

\section*{Acknowledgments}
\noindent The authors acknowledge financial support from LCNano/SisNANO 2.0 (grant 442591/2019$-$5), INCT $-$ Carbon Nanomaterials, INCT $-$ Materials Informatics, and INCT $-$ NanoVIDA. L.B. (grant E$-$26/202.091/2022 process 277806), O.M. (grant E$-$26/200.729/2023 process 285493)  and G.C. (grant E$-$26/200.627/2022 and E$-$26/210.391/2022 process 271814) are grateful for financial support from FAPERJ. The authors also acknowledge the computational support of N\'{u}cleo Avan\c{c}ado de Computa\c{c}\~{a}o de Alto Desempenho (NACAD/COPPE/UFRJ), Sistema Nacional de Processamento de Alto Desempenho (SINAPAD), Centro Nacional de Processamento de Alto Desempenho em S\~{a}o Paulo (CENAPAD$-$SP), and technical support of SMMOL$-$solutions in functionalyzed materials.

\section{Data availability}
\noindent Data will be made available on request.

\bibliographystyle{elsarticle-num}
\bibliography{tmm-bib}

\end{document}